\documentclass[11pt]{article}

\usepackage{amsmath,amssymb}
\usepackage{slashed}
\usepackage{xcolor} 
\usepackage{graphicx}
\usepackage{dcolumn} 
\usepackage{bm} 
\usepackage{epsfig}
\usepackage{epstopdf}
\usepackage{grffile}
\usepackage{color}
\usepackage{colordvi}
\usepackage{amsmath,amssymb}
\usepackage{rotating}
\usepackage{lscape}
\usepackage{cite}
\usepackage{float}
\usepackage{hyperref}

\usepackage{amsmath}
\usepackage{amssymb}
\usepackage{physics}
\usepackage{caption}

\newcommand{\beq}{\begin{equation}}
\newcommand{\eeq}{\end{equation}}

\def\Re{{\cal R \mskip-4mu \lower.1ex \hbox{\it e}\,}}
\def\Im{{\cal I \mskip-5mu \lower.1ex \hbox{\it m}\,}}

\def\tev{\,{\ifmmode\mathrm {TeV}\else TeV\fi}}
\def\gev{\,{\ifmmode\mathrm {GeV}\else GeV\fi}}
\def\mev{\,{\ifmmode\mathrm {MeV}\else MeV\fi}}
\def\to{\rightarrow}

\begin{document}

\begin{center}

\vspace*{15mm}
\vspace{1cm}
{\Large \bf New Probes for Axion-like Particles at Hadron Colliders}

\vspace{1cm}

\small{\bf Javad Ebadi$^{\dagger}$\footnote{javadebadi@ipm.ir},   Sara Khatibi$^{\bigstar,\ddagger}$\footnote{sara.khatibi@ut.ac.ir},  
Mojtaba Mohammadi Najafabadi$^{\ddagger}$\footnote{mojtaba@ipm.ir} }

 \vspace*{0.5cm}

{\small\sl 
$^{\dagger}$ School of Physics, Institute for Research in Fundamental Sciences, P.O. Box 19395-5531, Tehran, Iran \\
$^{\bigstar}$Department of Physics, University of Tehran, North Karegar Ave., Tehran 14395-547, Iran\\
$^{\ddagger}$School of Particles and Accelerators, Institute for Research in Fundamental Sciences (IPM) P.O. Box 19395-5531, Tehran, Iran \\
}

\vspace*{.2cm}
\end{center}

\vspace*{10mm}

%
%
\begin{abstract}\label{abstract}
Axion-like particles (ALPs) appear from spontaneous global symmetry
breaking in many extensions of the Standard Model (SM). 
In this paper, we find bounds on ALP ($a$) model parameters at the LHC from 
the ALP production associated with a photon and a jet ($j+\gamma+a$) as well as 
single top and top quark pairs ($t+j+a$, $t\bar{t}+a$) in a model independent approach.
In particular, it is shown that the ALP production associated with a photon plus a jet at the LHC is a promising 
channel with significant sensitivity to probe the ALP couplings to gluons and electroweak gauge bosons. 
The prospects are presented at the High Luminosity LHC including a realistic detector simulation and pile up effects.
Furthermore, the ALP model is examined through its contributions to the top quark (chromo)magnetic dipole moments. 
It is shown that the top quark magnetic and chromomagnetic dipole moments enable us to probe the ALP couplings to top
quark and gauge bosons at a time.
The constraints are complementary to those obtained from direct searches, as they are
sensitive to light ALPs.
\end{abstract}

\newpage

\section{Introduction}\label{sec:intro}

In spite of the remarkable achievements of the Standard Model (SM) of particles physics, 
there are observational and  theoretical aspects which are not completely understood or explained within the SM framework. 
Existence of Dark Matter (DM), neutrino mass, and baryon asymmetry are examples of observational problems.
The strong CP  and hierarchy problems are among the theoretical problems not addressed in the SM. 
So far, many theories beyond the Standard Model (BSM) have been constructed to explain these defects.
Although numerous attempts have already been done to figure out a footprint of these BSMs at the LHC,
no significant sign of new physics at high energies has been discovered yet. As a result, looking for new light degrees of freedom
or weakly coupled states to the SM content are taken into consideration these days.

Many new models predict one or more new light pseudoscalar particles in their spectrum which enable us to 
explain part of the SM shortcomings.
For instance, in order to solve the strong CP problem \cite{tHooft:1976rip,Dine:2000cj,Hook:2018dlk}, 
Peccei and Quinn \cite{Peccei:1977hh} proposed a mechanism with a spontaneously broken 
global $U(1)_{\text{PQ}}$ symmetry, which predicts a pseudo Nambu Goldstone boson, called QCD axion. 
Furthermore, the pseudoscalar particles can appear in other models, such as DM portal models \cite{Dolan:2014ska}, 
low-energy effective field theories of string theory \cite{Patrignani:2016xqp}
and some models which explain the muon magnetic moment  anomaly \cite{Chang:2000ii}.
In general, any model with global $U(1)$ symmetries, which are spontaneously broken, predicts 
pseudo Nambu Goldstone bosons so that their masses and couplings are independent parameters.
The pseudo Nambu Goldstone bosons in such models are called axion-like particles (ALPs).
The strength of the
couplings of the ALPs to SM matter fields is proportional to the inverse of the scale 
of $U(1)$ spontaneous symmetry breaking $f_{a}$, which is much larger than the SM electroweak symmetry breaking scale.
The ALP model is able to solve the observed matter-antimatter asymmetry for
the case that $f_{a}$ resides between about 100 TeV and $10^{4}$ TeV \cite{Jeong:2018jqe}. 
Pseudo-Nambu-Goldstone bosons could also appear in 
supersymmetric (SUSY) models with SUSY dynamical breaking and 
a spontaneously broken R-symmetry.  Such pseudo-Nambu-Goldstone bosons 
are called R-axion and naturally expected to be light \cite{Bellazzini:2017neg, Arganda:2018cuz}.

A considerable region of the parameter space of the ALP model has already been probed 
by  cosmological observations and low-energy experiments. 
Also, the LEP and LHC data have been employed to constrain the parameter space of ALP. 
If an  ALP is produced in a collision at high energy colliders, it has different properties
which can be classified into four categories.
First, ALP can be  long-lived  compared to the detector scales, 
therefore it can escape detection which results in a large missing energy  signature in the detectors.
We note that ALPs have a negligible interaction with detector material due to very stringent bounds on 
their interactions with fermions \cite{Brivio:2017ije}.
Second, ALP can decay to two massless gauge bosons inside the detector, 
and the detector could discriminate these two objects. 
The signature would contain two jets or two photons which are nearly back-to-back. 
This signature is expected if the ALP is heavy and
the extreme limit of this case could happen when the ALP is produced nearly at rest.
Third, ALP can be highly boosted and decays to two massless gauge bosons inside the detector, 
but is recognized as one object in the detector. 
In the case of decay of ALP into two gluons, the signature would be a "fat jet". 
For the photons, there is a possibility that detector recognize two photons as a single photon.
Fourth, the ALP decays into a $Z$ boson and a photon or into two massive gauge bosons $VV$ ($VV = ZZ, WW$).
In $a\rightarrow Z\gamma$, the partial decay width for light ALPs is suppressed by the $(m_{a}/m_{Z})^{4}$ factor.
More suppression for the case of $a \rightarrow VV$ is expected due to more limited phase space.

Decays  $Z\rightarrow \gamma \gamma$ and $Z\rightarrow \gamma \gamma \gamma$ provides
the possibility to exclude part of the parameter space of the ALP model
which was not excluded by low energy experiments \cite{Jaeckel:2015jla,Mimasu:2014nea}. 
Higgs exotic decay modes enable us to access part of the parameter space which could not be constrained by $Z$ boson decays.
The $h\rightarrow Z a \rightarrow \ell^+ \ell^- \gamma \gamma$ 
and $h \rightarrow a a \rightarrow \gamma\gamma\gamma\gamma$
channels at LHC are utilized to probe the ALP coupling to  the photon in terms of  the ALP mass \cite{Bauer:2017nlg,Bauer:2017ris}.
Moreover, there are studies based on large missing energy signature 
where the ALP escape detections and appears as missing momentum.
The mono-jet, mono-$\gamma$, dijet, and diphoton searches at the LHC and future colliders are employed
to constrain the ALP parameter space \cite{Mimasu:2014nea, Mariotti:2017vtv}. 
In addition, the mono-$W$, mono-$Z$, mono-Higgs channels and the associated production of ALP with $W+\gamma$ are 
also investigated in Ref.\cite{Brivio:2017ije}. 
There are other proposed ways to look for ALP at the colliders.
For example, the heavy ion ultra-peripheral collisions (UPCs) are utilized 
to search for axion coupling to photons \cite{Knapen:2016moh, Sirunyan:2018fhl, Bruce:2018yzs}.
Moreover, the ALP model parameter space could be probed in forward physics \cite{Baldenegro:2018hng}.  
Flavor factories using $B$ meson decays and flavor changing processes are also power tools to probe
the ALP couplings \cite{CidVidal:2018blh, Gavela:2019wzg}.
The ALP coupling with gluon has been studied based on a data-driven method 
for a region of ALP mass close to QCD scale to 3 GeV in Ref.\cite{yotam}. 

In this work, we study the associated production of an ALP with a photon and a jet,
single top quark plus an ALP, and $t\bar{t}$ production with an ALP at the LHC
to probe the ALP model parameter space.
This study concentrates on a region of the ALP parameter space,
in which it would not decay inside the detector and appears as missing transverse energy  ($E_T^{\text{miss}}$).
We also use the top quark (chromo)magnetic dipole moment
as a tool to limit the ALP couplings.

The organization of the paper is as follows. In section \ref{sec:framework}, 
a short introduction of the theoretical framework of the ALP model is given.
Section \ref{LHCsearch}  presents the colliders searches 
for the ALP model using the associated production of an ALP with
a jet and a photon as well as top quark(s).  In section \ref{dipolet},
the contributions of the ALP model to the top quark magnetic and chromomagnetic dipole moments
are calculated. Then, the upper limits on the (chromo)magnetic dipole moments are
used to constrain the model parameters. Finally, a summary of the results and conclusions are 
given in section \ref{con}.

%
\section{Theoretical framework}\label{sec:framework}
In this work,  an ALP is studied in a scenario where it is a singlet scalar under the SM 
gauge group and is odd under the CP transformation. 
The most general Lagrangian up to dimension $D=5$ operators
which describes the effective interactions of the ALP and SM fields is given by~\cite{Brivio:2017ije}:
\begin{eqnarray}
\begin{split}
	\mathcal{L}_{eff}^{D\leq5} 
	&= 
	\mathcal{L}_{SM} + \frac{1}{2} (\partial^\mu a)(\partial_\mu a) - \frac{1}{2} m_a^2 a^2\\
	&+ c_{a \Phi} \frac{\partial^\mu a}{f_a} (\Phi^\dagger i \overleftrightarrow{D}_\mu \Phi) 
	+ 	\frac{\partial^\mu a}{f_a} 
	\sum_{F} \bar{\Psi}_F \mathbf{C}_{F} \Psi_{F} \\
	&-  c_{GG} \frac{a}{f_a} G^{A}_{\mu\nu} \tilde{G}^{\mu\nu,a} 
	-  c_{BB} \frac{a}{f_a} B_{\mu\nu} \tilde{B}^{\mu\nu}
	-  c_{WW} \frac{a}{f_a} W^{a}_{\mu\nu} \tilde{W}^{\mu\nu,a},
\end{split}
\label{general-NLOLag-lin}
\end{eqnarray}
here the summation  is performed over all the SM fermions field 
${F = L_L, Q_L , e_R , d_R , u_R}$, where  $L_L$ and $Q_L$ are the $SU(2)_L$ doublets 
 and $e_R$,$d_R$,$u_R$ are $SU(2)_L$ singlet.  
 The $\mathbf{C}_F$ matrices are $3\times3$ 
 Hermitian matrices in flavor space. 
 The Higgs boson doublet is denoted by $\Phi$ and ALP  field is represented by $a$. 
 The $G_{\mu\nu}$, $W_{\mu\nu}$, and $B_{\mu\nu}$ are field strengths for $SU(3)_c$, $SU(2)_L$ and $U(1)_Y$,
 respectively. The $\tilde{X}^{\mu\nu}$ is defined as follow
\begin{equation}
	\tilde{X}^{\mu\nu} \equiv 
	\frac{1}{2} \epsilon^{\mu \nu\alpha \beta } X_{\alpha \beta},\nonumber
\end{equation}\label{equ::def::Xtilde}
where $\epsilon^{\alpha \beta \mu \nu} $ is the Levi-Civita symbol.
We note that the Lagrangian of Eq.\ref{general-NLOLag-lin} does not add any new source of CP violation 
other than what already exists in the SM. This Lagrangian respects 
gauge symmetry of SM and it is invariant under the CP transformation. 
In addition, neglecting the ALP mass term, 
the Lagrangian is invariant under shift transformation  up to total derivative terms
which is a manifestation of a broken global symmetry.
In this work, for simplicity we only focus on bosonic 
Lagrangian or in other words,  $\mathbf{C}_F=0$ are assumed for all five type of SM fermions.
 Then, after performing the following field redefinition:
\begin{eqnarray}
\Phi\to e^{ic_{a \Phi}\, a/f_a}\Phi,
\label{redef_OPhia}
\end{eqnarray}
the $\frac{\partial^\mu a}{f_{a}} (\Phi^\dagger i \overleftrightarrow{D}_\mu \Phi)$ operator 
is eliminated which causes the appearance shift-symmetry breaking terms in the Lagrangian. 
After the field redefinition Eq.\ref{redef_OPhia}, the Lagrangian up to dimension five operators  takes the following form \cite{Brivio:2017ije}:
\begin{eqnarray}
\begin{split}
	\mathcal{L}_{eff}^{D\leq5} 
	&= 
	\mathcal{L}_{SM} + 
	\frac{1}{2} (\partial^\mu a)(\partial_\mu a)
	- \frac{1}{2} m_a^2 a^2
	+ \\
	&-  c_{GG} \frac{a}{f_a} G^{A}_{\mu\nu} \tilde{G}^{\mu\nu,A} 
	-  c_{WW} \frac{a}{f_a} W^{A}_{\mu\nu} \tilde{W}^{\mu\nu,A}
	-  c_{BB} \frac{a}{f_a} B_{\mu\nu} \tilde{B}^{\mu\nu} 
	+ c_{a \Phi} \mathbf{O}^\psi_{a\Phi},
\end{split}
\label{bosonic_Lagrangian}
\end{eqnarray}
where
\beq
\mathbf{O}^\psi_{a\Phi}
\equiv 
i\left(\bar{Q}_L \mathbf{Y}_U\tilde\Phi u_R-\bar{Q}_L \mathbf{Y}_D\Phi d_R-\bar{L}_L\mathbf{Y}_E\Phi e_R\right)\frac{a}{f_a} + \text{h.c.}
\label{ALP-Yukawa}
\eeq
The higher order terms, which appear due to the field redefinition, have been omitted in the Lagrangian. 
\\
The Lagrangian of Eq.\ref{bosonic_Lagrangian} has been implemented in  \texttt{FeynRules} \cite{Alloul:2013bka} 
based on the study and notation of Ref.\cite{Brivio:2017ije}. 
Then, the obtained Universal FeynRules Output (UFO) \cite{Degrande:2011ua}\footnote{\url{http://feynrules.irmp.ucl.ac.be/attachment/wiki/ALPsEFT/ALP_linear_UFO.tar.gz}}  
model is inserted to \texttt{MadGraph5\_aMC@NLO}~\cite{Alwall:2011uj} 
to perform the numerical calculations of the cross sections  and to generate events.

\subsection{ALP decay} \label{subsec:Decay}

ALP can decay into charged leptons, photons, jets (gluons, quarks) or light hadrons
according to its mass and couplings to these particles.
The Lagrangian describing the decay of ALP to two photons and two gluons
can be written as follows:
\begin{eqnarray}
	\mathcal{L}_{eff}^{D\leq5}
	\supset
	-  c_{\gamma\gamma} \frac{a}{f_a} F_{\mu\nu} \tilde{F}^{\mu\nu}  -  c_{GG} \frac{a}{f_a} G^{A}_{\mu\nu} \tilde{G}^{\mu\nu,A}, 
\label{equ::photon-part-of-Lagrangian}
\end{eqnarray}
where $F_{\mu\nu}$ and $G^{A}_{\mu\nu}$ are the field strengths of the photon and gluon, respectively.  $c_{GG} $ and $c_{\gamma\gamma}$
are the couplings of gluon and photon with the ALP.  The coupling with photon $c_{\gamma\gamma}$ 
is related to $c_{BB}$ and $c_{WW}$ of Eq.\ref{bosonic_Lagrangian} via:
\begin{eqnarray}
	c_{\gamma\gamma} = \frac{1}{f_a} (c_{BB} \cos\theta_W^2 + c_{WW} \sin\theta_W^2),
	\label{equ::coupling-of-ALP-to-photons}
\end{eqnarray}
where $\theta_W$ is the Weinberg angle.  
There are several studies to constrain the ALP mass and its coupling to photons $c_{\gamma\gamma}$ 
obtained from various experiments ranging from low energy experiments to 
high energy colliders and cosmological observations \cite{Bauer:2017nlg,Bauer:2017ris}. \\
The decay rates of an ALP into two photons and two gluons at leading order
can be written as:
\begin{equation}
\Gamma_{a \rightarrow \gamma\gamma} =\Bigl(\frac{c_{\gamma\gamma}}{f_a}\Bigr)^2 \frac{m_a^3}{4\pi}~, ~ \\
 \Gamma_{a \rightarrow g g} =8\times\Bigl(\frac{c_{GG}}{f_a}\Bigr)^2 \frac{m_a^3}{4\pi}.  
\end{equation} 
where the decay rate of $a \rightarrow g g$ is calculated using pQCD which only is valid providing that $m_{a} \gg \Lambda_{\text{QCD}}$.
Chiral perturbation theory can be used for studying the interaction of gluon with ALP  for the case of  $m_{a} \lesssim 1$ GeV.
In Ref.\cite{yotam}, a data-driven method has been proposed to determine $c_{GG}$
for the region of $m_{a}$ close to QCD scale to 3 GeV using inclusive decays of $b\rightarrow s+a$,  and $\phi$, $\eta'$,
 $B^{\pm}$, and $B^{0}$ decays.

\section{Associated production of an ALP at the LHC}\label{LHCsearch}

In this section, we concentrate on constraining the ALP model parameters
through various processes with ALP in the final state in proton-proton collisions at the LHC. 
In particular, $j+\gamma+a$, $ t+j+a$, and 
$t\bar{t}+a$ are studied considering a realistic detector simulation including the main
background processes.

\subsection{ALP production with a photon and a jet}\label{sub:AphotonJet}

\begin{figure}
\begin{center}
	\resizebox{0.45\textwidth}{!}{\includegraphics{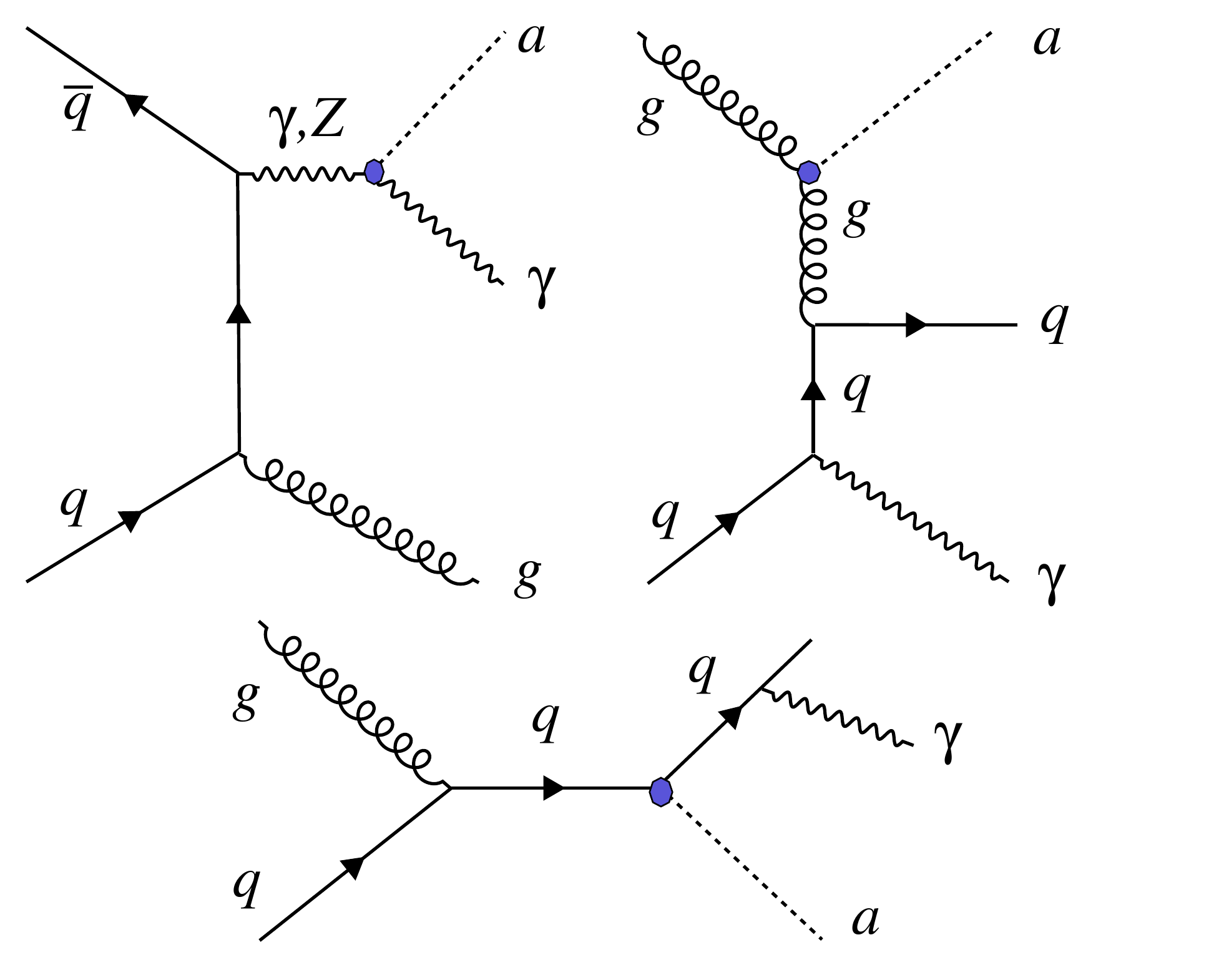}}
	\caption{Representative Feynman diagrams for production of an axion-like particle in association with a jet and a photon in 
	 proton-proton collisions at the LHC.}\label{Feynmanja}
	\end{center}
\end{figure}

We examine the potential of an ALP production associated with
a photon and a jet to probe the parameter space of the model at the LHC.
Although ALP is not directly detected by
the LHC detectors, its production could be deduced from observation of events with an
imbalance in the transverse momentum. \\
The production of an ALP associated with a photon and jet that 
has a final state of a photon, a jet, and missing transverse momentum
is identifiable with large efficiency and purity.
Figure \ref{Feynmanja} shows the representative Feynman diagrams
for $j+\gamma+a$ in proton-proton collisions at the LHC. 
One of the interesting features of this process is its sensitivity to all couplings 
of ALP, {\it i.e.} $c_{WW}, c_{BB},c_{GG},$ and $c_{a\Phi}$.

\begin{figure}
	\centering
\resizebox{0.55\textwidth}{!}{\includegraphics{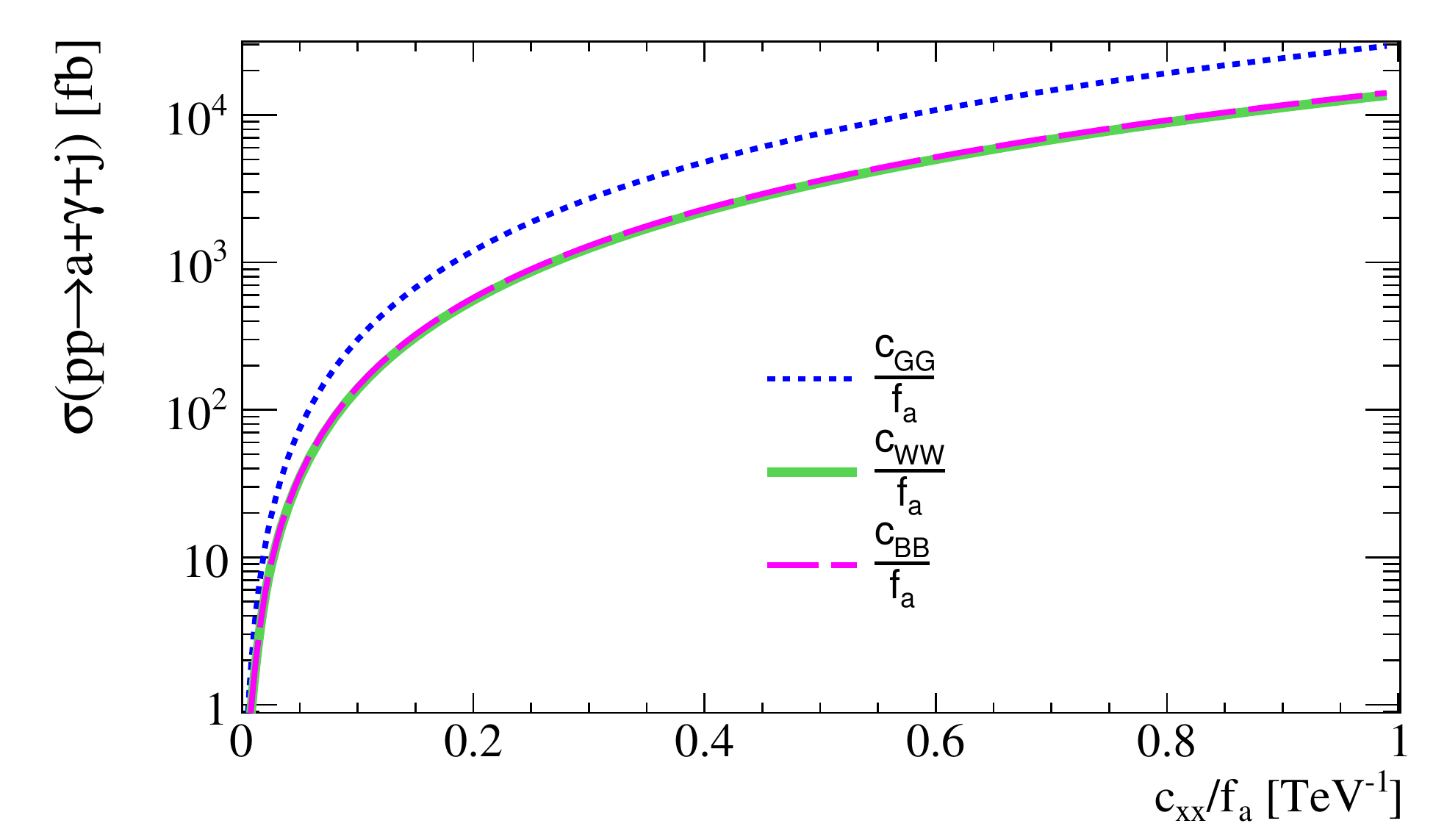} }
	\caption{Production cross section of $j+\gamma+a$ as a function of couplings $c_{GG}$, $c_{BB}$, $c_{WW}$ at the LHC with the
	center of mass energy of 14 TeV.}
	\label{fig:crossSection}
\end{figure}

Assuming a single non-vanishing ALP coupling 
at a time, the leading order (LO) cross sections $\sigma(pp \rightarrow j+\gamma+a)(c_{XX})$
read:
\begin{eqnarray}
&&    \sigma({c_{GG}}) = 30.0  \Bigl(\frac{c_{GG}}{f_a}\Bigr)^2 \text{pb},  ~
    \sigma({c_{WW}} )=13.9  \Bigl(\frac{c_{WW}}{f_a}\Bigr)^2  \text{pb}, \nonumber \\  
&&    \sigma({c_{BB}}) = 14.4  \Bigl(\frac{c_{BB}}{f_a}\Bigr)^2  \text{pb}, ~
    \sigma({c_{a\Phi}}) = 0.36  \Bigl(\frac{c_{a\Phi}}{f_a }\Bigr)^2  \text{pb},
    \label{cs}
\end{eqnarray}
where $f_{a}$ is in TeV unit. The cross sections are obtained with \texttt{MadGraph5\_aMC@NLO}
using the NNPDF23 \cite{Ball:2012cx} as the proton parton distribution function (PDF).  
The cross sections are calculated for the value of ALP mass $m_{a} = 1$ MeV and 
are found to be almost independent of $m_{a}$ 
for any value of $m_a \lesssim ~\text{few GeV}$
which is expected due to the fact that 
$m_{a}$ is negligible comparing to the typical energy scale of the $j+\gamma+a$ process.
The dependence of the rate of $j+\gamma+a$ 
on the couplings of ALP to gluon and weak gauge bosons are shown in Fig.\ref{fig:crossSection}.  
For large values of $m_a$ the cross section decreases because of the low probability of production 
of heavy particle in final state. Among various processes in $j+\gamma+a$ production, those
with a gluon and a quark in the initial state has larger contribution to the cross section than those from quark-antiquark annihilation.
This is because of the large gluon PDF in particular at low $x$  where 
the process mostly occurs.
As it can be seen in Eq.\ref{cs}, there is more sensitivity to the ALP coupling to gluon, {\it i.e.} $c_{GG}$,
which is due to the fact that it appears in processes with both initial states $gq(\bar{q})$ and $q\bar{q}$ 
as well as the large gluon PDF in the $gq(\bar{q})$ part. 
As it is obvious, the cross section has a negligible  dependence on $c_{a\Phi}$ coupling,  $\sigma \sim \order{10^{-1}} \text{fb}$. Therefore, 
in this work the effect of $\mathbf{O}^\psi_{a\Phi}$ operator in $j+\gamma+a$ channel is not studied. 
Using the Beam Dump experiments, the linear combination of  $c_{WW}$ and $c_{BB}$ has been measured to be very small \cite{Brivio:2017ije}: 
\begin{eqnarray}
|\frac{c_{BB}}{f_{a}} \cos\theta_W^2 + \frac{c_{WW}}{f_{a}} \sin\theta_W^2| \leq ~ 2.5\times 10^{-3} ~\text{for } ~m_{a} \leq 1 ~\text{MeV}. 
\end{eqnarray} 
Therefore, in this work we skip $c_{BB}/f_{a}$ coupling and we obtain
the limit on $c_{BB}/f_{a}$ using the limit on $c_{WW}/f_{a}$.

The main background contributions to signal ($j+\gamma+a$) arise from the following
sources:
\begin{itemize}
\item{Irreducible background from $Z\gamma+j$ events, where $Z$ boson decays to invisible neutrinos $Z\rightarrow \nu\bar{\nu}$.  }
\item{$W\gamma+j$, $t\bar{t}\gamma$, and $t\gamma j$ processes with 
the $W$ boson (directly produced or from the top quark decay) decays 
into $l (= e,\mu,\tau)+\nu$, where $l$ is out of detector acceptance. In these processes, neutrino(s) from $W$ boson decays
is a source of genuine missing transverse momentum.}
\item{$\gamma+j$ events where missing transverse momentum arises from leptonic decays
of hadrons inside the jets, mis-measurement of jet energy, or detector noise. $W\gamma+j$, $t\bar{t}\gamma$, and $t\gamma j$
with hadronic decays of the $W$ boson and top quark(s) are also of this type. }
\end{itemize} 
 
 The $Z\gamma+j$ and $W\gamma+j$ backgrounds are estimated based on a data-driven technique
 while the rest are calculated from simulation.
Now, we turn to the simulation of the signal and SM
background processes.  The signal and background events are generated with \texttt{MadGraph5\_aMC@NLO}
and passed through \texttt{Pythia}~\cite{Sjostrand:2006za} for showering, hadronization, and decays of unstable particles. 
Then, the events are passed to \texttt{Delphes}~\cite{deFavereau:2013fsa} 
for simulation of an upgraded CMS detector \cite{Chatrchyan:2008aa, CMSCollaboration:2015zni}
including additional proton-proton interactions per bunch crossing (pileup) with an average of 200. 
The jet finding is performed via \texttt{FastJet}~\cite{Cacciari:2011ma}.
The anti-$k_t$ algorithm is used for jet reconstruction with a distance parameter of 0.4 \cite{Cacciari:2008gp}
including pileup correction.
To generate signal events, we consider the effect of one coupling at a time and
various samples with ALP masses from 1 MeV to 300 MeV for $c_{GG}$ are generated and $f_a$ is set to $1~\text{TeV}$.
To find the exclusion regions in the parameter space $(c_{GG},m_{a})$, the signal selection efficiencies are obtained 
as a function of $c_{GG}$  and $m_{a}$.
For simplicity, the search for the sensitivity on $c_{WW}$
 coupling is performed only for one ALP mass benchmark of $m_{a} = 1$ MeV. 

It should be noted that a fraction of ALPs decay inside the detector volume which would not appear 
as missing energy.  The  ALP decay length $L_{a}$  is proportional to $ \sqrt{\gamma^{2}-1}/\Gamma_{a}$, where 
$\gamma$ and $\Gamma_{a}$ are the ALP Lorentz factor in each event and its total width, respectively. 
The probability that the ALP decays in the detector is proportional to $e^{-L_{\text{det}}/L_{a}}$,  
where $L_{\text{det}}$ is the distance from the collision point to the detector component
in which the ALP is reconstructed. In this analysis, the probability that the ALP escapes the detector
is considered event-by-event. 

Events are selected by requiring exactly one isolated
photon with a transverse
momentum $p_{\rm T} \geq $ 20 GeV and a pseudorapidity
$|\eta| \leq 3.0$. Photon isolation is applied using a pileup corrected isolation
variable $I_{rel}$ as defined in Ref.\cite{deFavereau:2013fsa} 
which assures negligible activity in the vicinity of the photon.
$I_{rel}$ is obtained from the amount of transverse energy $p_{\rm T}$, calculated relatively
to the photon $p_{\rm T}$, in a cone of radius $R = \sqrt{\eta^{2}+\phi^{2}} = 0.3$
around the photon candidate, where
$\phi$ is the azimuthal angle with respect to the $z$ direction.
The selected  photon is required to satisfy $I_{rel} < 0.15$.
Also, it is required to have at least one jet with $p_{\rm T} \geq 20$ GeV and $|\eta| \leq 4.0$. In order to 
suppress backgrounds with high jet multiplicity such as $t\bar{t}\gamma$ the number of jets is required to 
be less than three with at most one b-jet. To have well-isolated objects, the angular separation 
between photon and jets, and between jets are required to be greater 
than 0.4, {\it i.e.} $\Delta R(i,j) = \sqrt{(\Delta\eta_{ij})^{2}+(\Delta\phi_{ij})^{2}} \ge 0.4$.
The minimum cut on missing transverse energy $E_T^\text{miss}$ is chosen so that the analysis sensitivity to
each ALP coupling is maximized. For the signal scenario of non-zero $c_{GG}$, the 
optimized cut on $E_T^\text{miss}$ is found to be $90$ GeV while 
for the $c_{WW}$ the best sensitivity is achievable with $E_T^\text{miss} \geq 50$ GeV.
Further $\gamma+j$ background suppression could be obtained by applying an upper cut on $\Delta\phi(j,\gamma)$. 
Events of $\gamma+j$ are expected to be back-to-back distributed mostly around $\Delta\phi(j,\gamma) \sim \pi$
while this not the case for signal due to the presence of an ALP in the final.  Therefore, an upper cut of 
2.7 is applied on $\Delta\phi(j,\gamma)$.

In order to ensure the validity of the effective Lagrangian, its suppression scale ($f_{a}$)
is required to be far above the typical energy scale of the process. As a result,  one must require that the
energy scale of the process $\sqrt{\hat{s}}$ to be much less than $f_{a}$ in each event.
In the processes under study in this work where ALP is in the final state appearing as missing energy, $\sqrt{\hat{s}}$ is not fully
measurable. Therefore, to ensure the validity of the effective theory, $f_{a}$ is compared to 
$E_T^\text{miss}$. In each event, it is required that $E_T^\text{miss}  <  f_{a}$.
The same approach is applied in the rest of the paper in other processes.

The production of a $Z$ boson in association with a photon and jets followed by invisible $Z$ boson decay is
an irreducible background. We estimate this background using a data-driven method which
relies on $Z(\rightarrow \mu^{+}\mu^{-})\gamma+j$ events then we compare the results with 
simulation.
In this technique, the pair of muons ($\mu^{+}\mu^{-}$) is interpreted 
as missing momentum in the $Z(\rightarrow \nu\bar{\nu})\gamma+j$ process keeping
selection criteria on the objects. The number of $Z(\rightarrow \nu\bar{\nu})\gamma+j$ events is
measured as the number of $Z(\rightarrow \mu^{+}\mu^{-})\gamma+j$ events corrected for the acceptance cuts efficiency ($A$), 
detector efficiencies $\epsilon$ and
the branching fraction ratio: 
\begin{eqnarray}
\text{Correction Factor} = \frac{Br(Z\rightarrow \nu\bar{\nu})}{Br(Z\rightarrow \mu^{+}\mu^{-})}\times A\times \epsilon,
\end{eqnarray}
where $Br(Z\rightarrow \nu\bar{\nu})$ and $Br(Z\rightarrow \mu^{+}\mu^{-})$ are the branching fractions of invisible and leptonic
decays of $Z$ boson.

\begin{figure}
\centering
	\resizebox{0.46\textwidth}{!}{\includegraphics{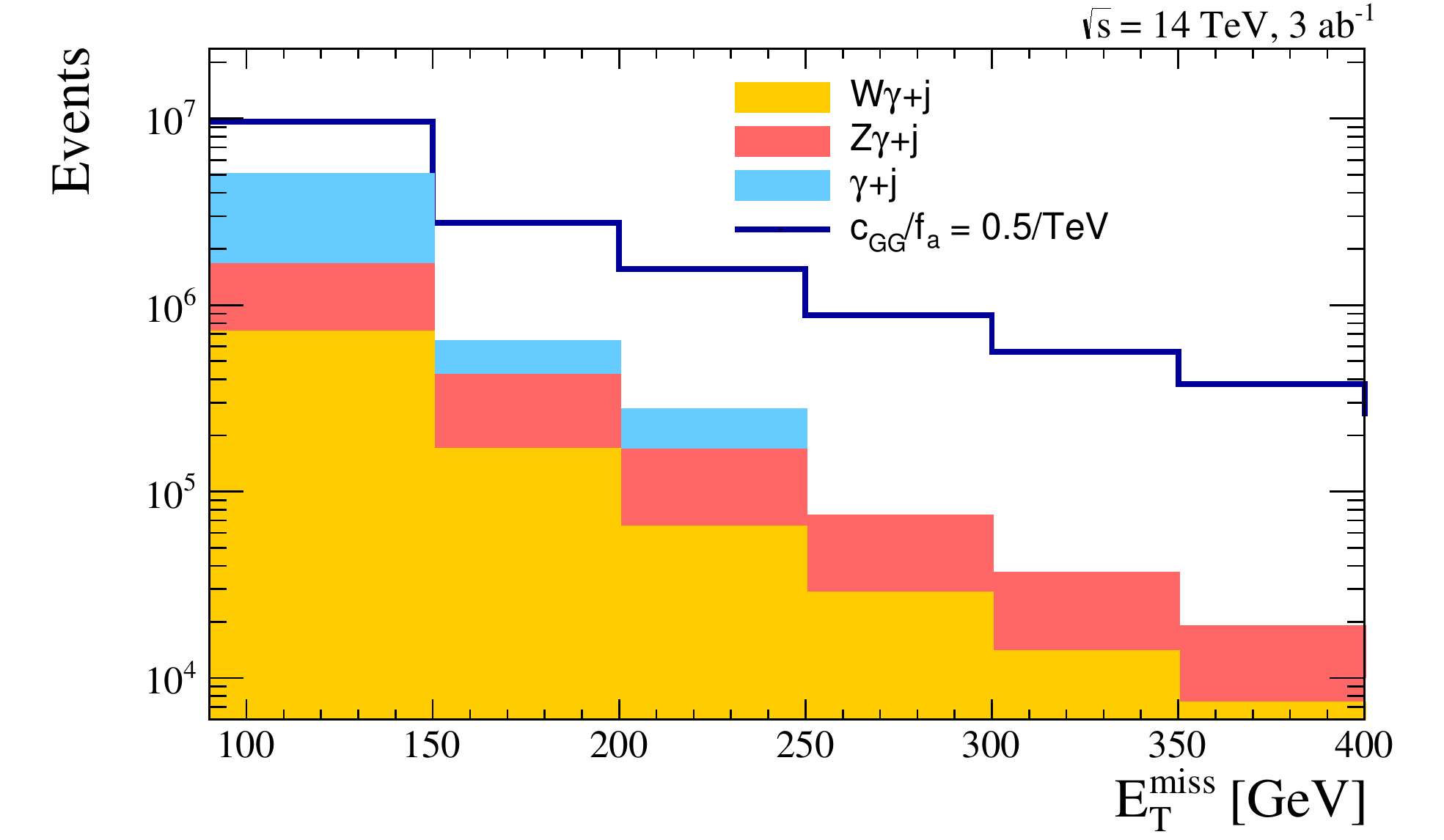}}
	\caption{ Distribution of $E_T^{\text{miss}}$ for signal ($c_{GG}/f_{a} = 0.5$ TeV$^{-1}$, $m_{a} = 1$ MeV) (black line) 
	  $W\gamma+j$ (orange), $Z\gamma+j$ (red), and $\gamma+j$ (blue) 
	   for an integrated luminosity of 3000 fb$^{-1}$ at the LHC with $\sqrt{s} = 14$ TeV.}\label{fig:deltaPhiMET}
\end{figure}

Similar technique is used for estimation of the $W\gamma+j$ background,
where $W$ boson decays leptonically. However, additional correction due to
different topology needs to be included.  The number of estimated background
using the above method is consistent with the prediction from simulation within
$1\%$.
\\
Fig.\ref{fig:deltaPhiMET} represents the distribution of $E_T^\text{miss}$ 
for signal with $c_{GG}/f_{a} = 0.5$ TeV$^{-1}$, $m_{a} = 1$ MeV and various backgrounds with an integrated luminosity
of 3000 fb$^{-1}$. 
Table \ref{Table:Cut-Table} shows the efficiencies for the signal scenarios of $c_{GG}/f_{a} = 0.5$ TeV$^{-1}$, $c_{WW}/f_{a} = 0.5$ TeV$^{-1}$,
and the main background processes
after applying different cuts. The total number of background events corresponding to 3000 fb$^{-1}$ after  
$E_T^\text{miss} \geq 90$ GeV is $6.02 \times 10^{6}$.

\begin{table}[ht]
	\begin{center}
		\begin{tabular}{|c|c|c|c|c|c|}  \hline 
 Cut       &   $c_{GG}/f_{a} = 0.5$  TeV$^{-1}$          &    $c_{WW}/f_{a} = 0.5$ TeV$^{-1}$  &  $Z\gamma+j$   & $W\gamma+j$ & $\gamma+j$    \\   \hline
Jet and Photon     &         0.786          &   0.890                &  0.846      &  0.370      &  0.717 \\ \hline  
$\Delta\phi(j,\gamma) \leq 2.7$  &        0.671            &    0.541             &  0.636     &    0.227    &   0.142   \\ \hline
$E_T^\text{miss} \geq 50$ GeV  &        0.606           &    {\bf 0.237}      &  0.493    &   0.113     &  $3.7\times 10^{-4}$   \\ \hline
$E_T^\text{miss} \geq 90$ GeV  &         {\bf 0.449}   &    0.04             &  0.265     &     0.044   &   $3.8\times 10^{-5}$ \\ \hline
		\end{tabular}
			\caption{Efficiency for the signal and dominant SM backgrounds. The signal scenarios are 
		corresponding to values of $c_{GG}/f_{a} = 0.5$ TeV$^{-1}$, $c_{WW}/f_{a} = 0.5$ TeV$^{-1}$ and $m_{a} = 1$ MeV.  }
		\label{Table:Cut-Table}
		\end{center}
\end{table}

\begin{figure}
\begin{center}
	\resizebox{0.65\textwidth}{!}{\includegraphics{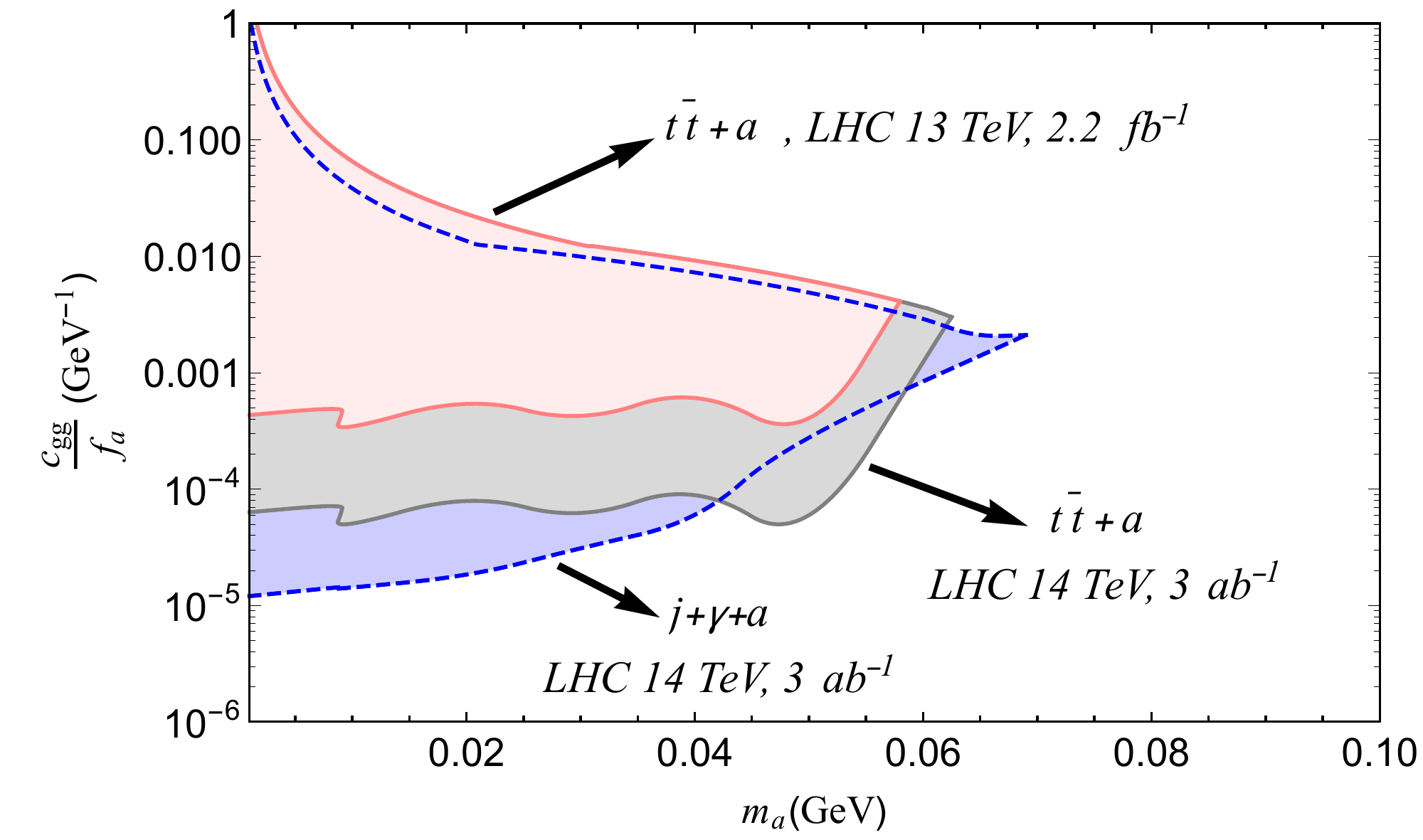}}
	\caption{The $95\%$ CL excluded regions of the parameter space of the ALP model $(c_{GG}/f_{a} ,  m_{a})$ derived from 
	$j+\gamma+a$ and $t\bar{t}+a$ channels. The  regions assuming a new result of the collider search using the $t\bar{t}+a$
	 process are shown with integrated luminosities of  2.2 fb$^{-1}$ and 3000 fb$^{-1}$ at the center-of-mass energies of 13 TeV and 14 TeV, respectively.  }
	    \label{fig:cgg-bound}
	\end{center}
\end{figure}

Constraints on the ALP coupling $c_{GG}/f_{a}$ and mass derived from 
$j+\gamma+a$ channel 
with 3 ab$^{-1}$ integrated luminosity of data is presented as blue-dashed region in Fig.\ref{fig:cgg-bound}.
As can be seen, better sensitivity for lighter ALPs is obtained.
The $95\%$ CL upper limits on $c_{GG}/f_{a}$ and $c_{WW}/f_{a}$ for $m_{a} = 1$ MeV with 3000 fb$^{-1}$
of data are found to be:
\begin{eqnarray}
\abs{c_{GG}/f_a} \leq 0.011 ~\text{TeV}^{-1}, ~ \abs{c_{WW}/f_a}  \leq 0.036~ \text{TeV}^{-1}.
\end{eqnarray}
The constraint on $c_{GG}/f_{a}$ at $95\%$ CL from mono-jet events from the LHC experiments using 19.6 fb$^{-1}$
of 8 TeV data is $0.025$ TeV$^{-1}$ \cite{Mimasu:2014nea}.  The $j+\gamma+a$ process  provides
comparable bound on $c_{GG}$ to the one from mono-jet events and it 
could be a complementary channel to the mono-jet channel.
The $95\%$ CL upper limits on $c_{WW}$ from the LHC at $\sqrt{s} = 13$ TeV with 
3000 fb$^{-1}$ from different channels are \cite{Mimasu:2014nea,Brivio:2017ije}:
\begin{eqnarray}
\abs{c_{WW}/f_a} & \leq& 0.14~ \text{TeV}^{-1}~\text{mono-photon},\nonumber \\
\abs{c_{WW}/f_a}  &\leq& 0.05~ \text{TeV}^{-1}~\text{mono-$Z$},\\
\abs{c_{WW}/f_a}  &\leq &0.16~ \text{TeV}^{-1}~\text{mono-$W$}.\nonumber
\end{eqnarray}
The bound obtained from mono-$W$ 
channel is the weakest  and the strongest limit is from the mono-$Z$ process.
One can see that the $j+\gamma+a$ process in this study provides  better sensitivity 
to $c_{WW}$ with respect to the other processes.

\subsection{ ALP production in association with a single top quark  }

\begin{figure}
\begin{center}
	\resizebox{0.45\textwidth}{!}{\includegraphics{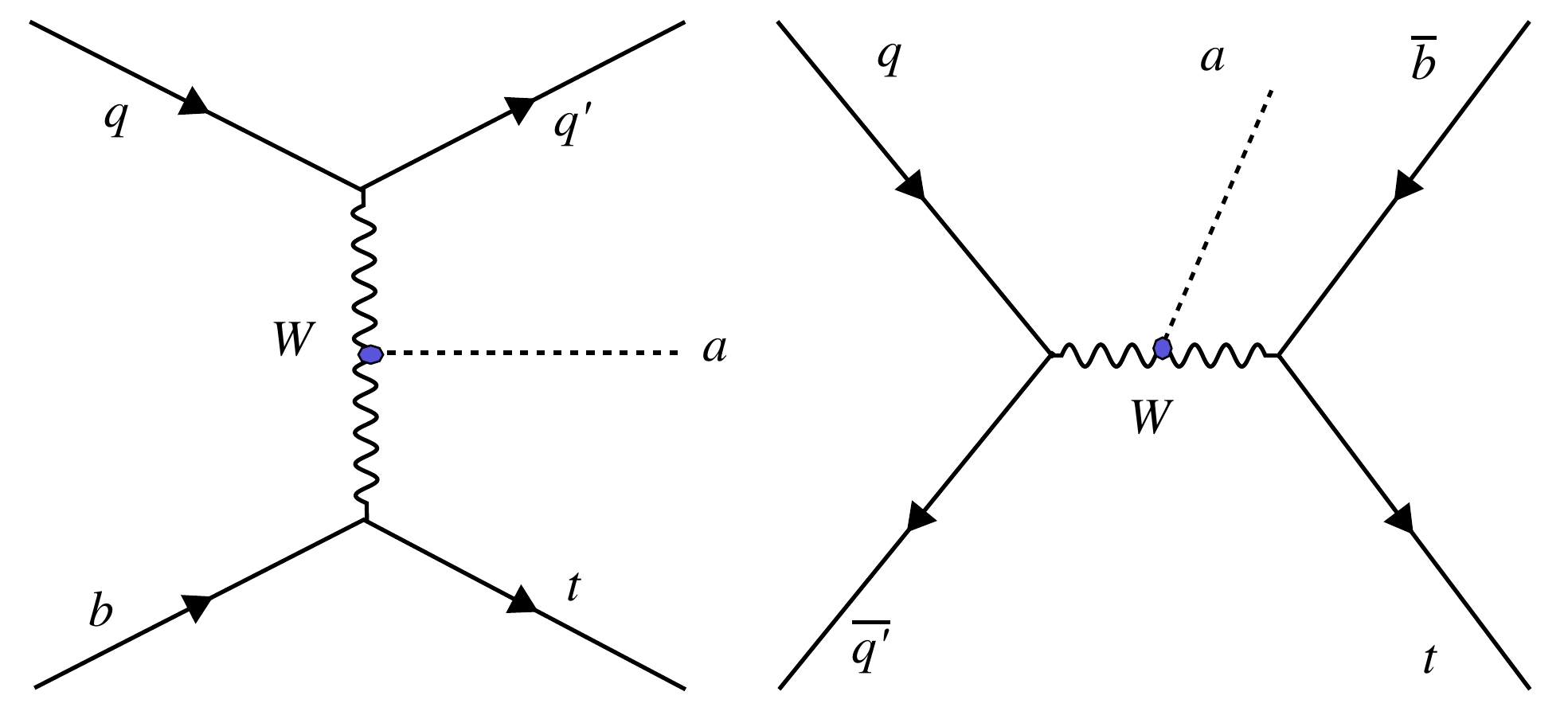}}
	\caption{Representative Feynman diagrams for  production of an ALP with a  top quark in the t- and s-channel modes at the LHC.}\label{Feynmantop}
	\end{center}
\end{figure}

 In this sub-section, we discuss the potential of the production of an ALP
 in association with a top quark at the LHC.
 Single top quark can be produced in association with 
 an ALP via three different mechanisms:
t-channel, s-channel, and tW production channel, where the t-channel has the 
largest cross section. We concentrate on studying the single top
quark production in association with an ALP in the
t- and s-channel. Figure \ref{Feynmantop} shows representative Feynman diagrams for this process.
The process is sensitive to $c_{WW}$ and has negligible sensitivity to $c_{a\Phi}$. The 
cross section dependence on $c_{WW}$, assuming $m_{a} = 1$ MeV,  has the following form:
\begin{eqnarray}
\sigma(pp \rightarrow t+j+a) = 0.12 \times (\frac{c_{WW}}{f_{a} })^{2} ~\text{pb}.
\end{eqnarray}

The search uses events with a top quark, at least one light-flavor jet, and missing transverse energy 
with the top quark decays into a bottom quark and a $W$ boson, followed by the
$W$ decay into a charged lepton ($e,\mu$) and a neutrino.
The final state consists of an electron or a muon, missing transverse energy, and at least two jets
from which one (or both) should originate from the hadronization of a b-quark.

The main backgrounds processes to the signal are $t\bar{t}$, SM single top processes (t-,s-, and tW-channel), $W+j$,
$Z+j$, and $WZ$ productions. The event generation, detector simulation, and reconstruction are
performed with similar tools and techniques as section \ref{sub:AphotonJet}. 
To select signal events, we require to have one isolated charged lepton ($e,\mu$)
with $p_{\rm T} \ge 20$ GeV and $|\eta|  \leq 3.0$. 
In addition, each event is required to have two or three jets  with $ p_{\rm T} \ge 20$ GeV
out of which at least one should be tagged as a b-jet with $|\eta| \leq 3.5$. 
To ensure all selected objects are well-isolated, the angular separation between all
objects should be $\Delta R  \ge 0.4$. 
The optimized lower cut on $E_T^\text{miss}$ is 250 GeV 
which reduces the  contribution of all SM background processes significantly.  
All the efficiency for the signal and SM backgrounds after final cuts are represented in Table \ref{xxxx}.
The total number of background events after all cuts with an integrated luminosity of 3000 fb$^{-1}$
is $1.41 \times 10^{6}$.

\begin{table}[htbp]
\begin{center}
\begin{tabular}{ |c|c|c|c|c|c|c| }
\hline
	 Signal ($c_{WW}/f_{a} =0.1~(0.5)$ TeV$^{-1}$)    &    $t+j$ & $t+W$ &  $t\bar{t}$   & $W+j$    & $Z+j$   &   $WZ$   \\
	\hline
	 $5.11\% ~(5.16\%)$    &    $0.014\%$   &   $0.02\%$    &    $0.01\%$    &   $0.02\%$  &    $0.0001\%$   &   $0.08\%$  \\
	\hline
\end{tabular}
\caption{Efficiency of signal with $c_{WW}/f_{a} = 0.1,0.5$ TeV$^{-1}$, $m_{a} = 1$ MeV and of different background processes after applying all cuts
are presented.}\label{xxxx}
\end{center}
\end{table}

The upper limit on $c_{WW}/f_{a}$ at $95\%$ CL for $m_{a} = 1$ MeV is obtained as:
\begin{eqnarray}
\abs{c_{WW}/f_a}  \leq  0.75  ~\text{TeV}^{-1}.
\end{eqnarray}
The limit is looser than the those obtained from $j+\gamma+a$ in section \ref{sub:AphotonJet}.
This was expected as the dependence of  $t+j+a$ cross section  to $c_{WW}/f_{a}$ is weaker than 
the cross section of $j+\gamma+a$ by two orders of magnitude.

\subsection{ ALP production in association with a pair of $t \bar{t}$ }

\begin{figure}
\begin{center}
	\resizebox{0.45\textwidth}{!}{\includegraphics{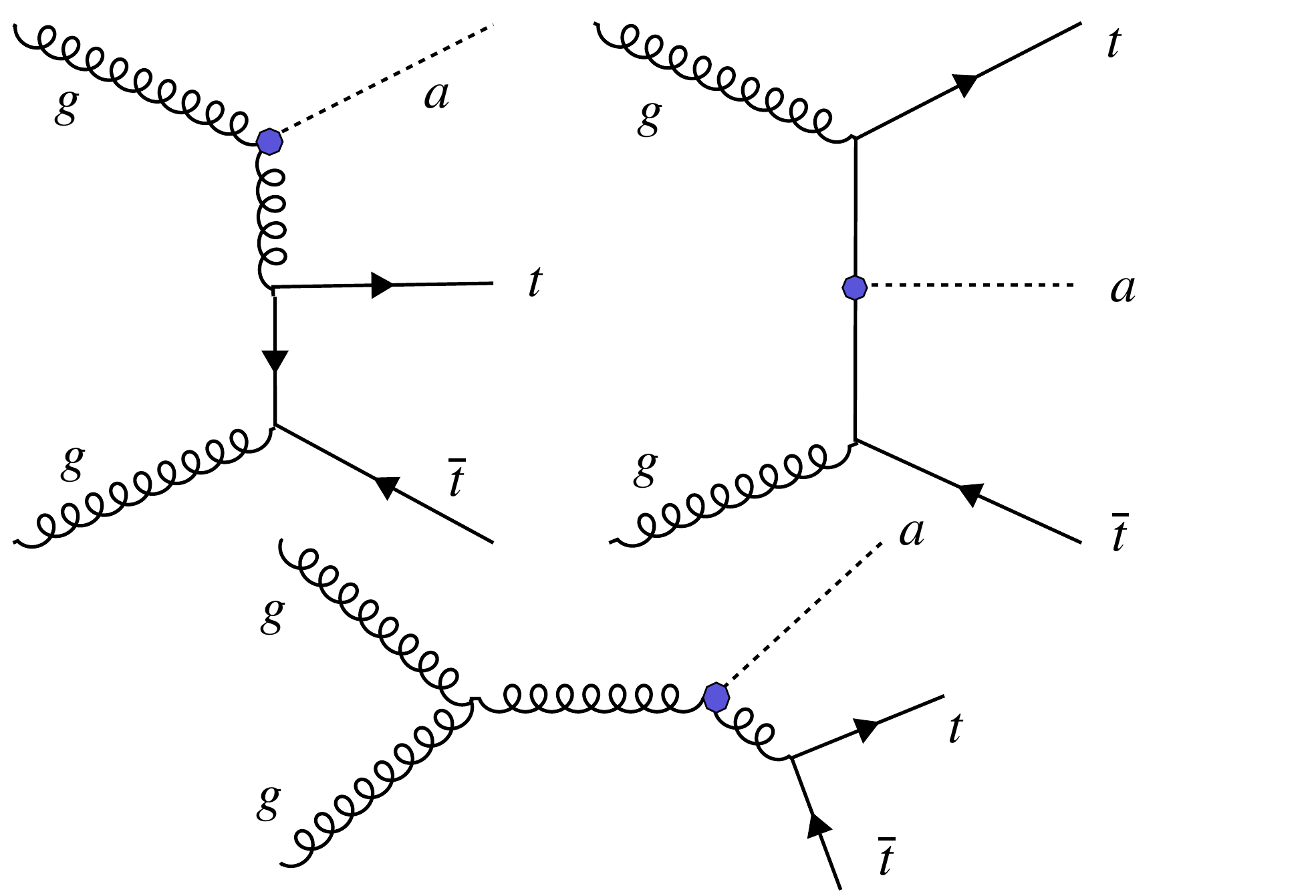}}
	\caption{Leading order Feynman diagrams describing the production of a pair of top quark with an ALP at the LHC.}\label{Feynman1top}
	\end{center}
\end{figure}

In this part, we investigate the potential of the top quark pair production in association
with an ALP at the LHC.  Representative leading order Feynman diagrams for $t\bar{t}+a$
production are depicted in Figure \ref{Feynman1top}. 
The process has a considerable sensitivity to $c_{GG}$ and has low sensitivity to 
$c_{a\Phi}$. The inclusive cross section up to order $f_{a}^{-2}$ for $m_{a}  = 1$ MeV can be written as:
\begin{eqnarray}
\sigma(pp \rightarrow t\bar{t}+a) = 3.9 \times (\frac{c_{GG}}{f_{a}})^{2}~\text{pb},~\sigma(pp \rightarrow t\bar{t}+a) = 0.04\times (\frac{c_{a\Phi}}{f_{a}})^{2}~\text{pb}.
\end{eqnarray}

At the LHC, the $t\bar{t}+a$ process can be probed directly via the $t\bar{t} + E_T^\text{miss}$ signature.
In Refs.\cite{Sirunyan:2017xgm, Aaboud:2017rzf}, the CMS and ATLAS collaborations performed searches for
simplified models for dark matter production assuming the existence of a mediator
that couples to both the dark matter and SM particles. Particularly, the concentration of these searches 
is on the case of production of a fermionic dark matter  via
 the exchange of a color-neutral scalar or pseudoscalar particle.
 In these models, the coupling between the new (pseudo)scalar 
state and SM particles is Yukawa-like as a result  proportional to the fermion masses.
It follows that the scalar mediator would be produced  mostly in association with heavy quarks
(in particular top quark) or via loop induced gluon fusion.
The characteristic signature to search for dark matter in $t\bar{t}+\phi (\rightarrow \chi\chi)$ channel
 is the presence of a high missing transverse momentum recoiling against 
$t\bar{t}$. This signature is similar to $t\bar{t}+a$. Therefore, we reinterpret 
the results of the CMS experiment search at the center of mass energy 13 TeV to the ALP model and extrapolate 
the results to larger integrated luminosity of data at the center of mass energy of 14 TeV.

We focus on the semi-leptonic $t\bar{t}$ channel and follow similar selection 
as applied the CMS experiment search \cite{Sirunyan:2017xgm}.
Events are selected by requiring 
one isolated lepton ($e,\mu$) with $p_{\rm T} \ge 30$ GeV and $|\eta| < 2.5$,
at least three jets from which one 
has to be a b-tagged jet. Jets are required to have $p_{\rm T} \ge 30$ GeV and 
$|\eta|  \leq 2.4$.  
Events with additional charged leptons with $p_{\rm T} \ge 10$ GeV that satisfy loose identification 
criteria are rejected. The magnitude of missing transverse momentum should be greater than 
160 GeV.
 For reducing the background contribution from 
$t\bar{t}$ and $W+j$, the transverse mass $M_{\rm T} = \sqrt{2p_{T,l}E_{T}^{\rm miss}(1-\cos\Delta\phi(\vec{p}_{T,l},\vec{E}_{T}^{\rm miss}))}$
is required to be larger than 160 GeV. 
The magnitude of the vector sum of jets with $p_{T} \ge 20~\text{GeV}$ and $\abs{\eta} \leq 5.0$, $\text{H}_\text{T}^\text{miss}$, 
is required to be greater than 120 GeV.
For further suppression of SM $t\bar{t}$, additional cut on $\text{M}_{\text{T}2}^\text{W}$,
introduced in Ref.\cite{Bai:2012gs}, is applied. It is required that   $\text{M}_{\text{T}2}^\text{W} \ge 200$ GeV.
The predicted number of background events corresponding to 2.2 fb$^{-1}$ is 43.2. 
The corresponding upper bounds on $c_{GG}/f_{a}$ with 
the integrated luminosities of 2.2 and 100 fb$^{-1}$ for $m_{a} = 1$ MeV are found to be:
\begin{eqnarray}
\frac{c_{GG}}{f_{a}} \le  0.44 ~\text{TeV}^{-1} ~ @ ~2.2 ~\text{fb}^{-1}  ~,~\frac{c_{GG}}{f_{a}} \le     0.15  ~\text{TeV}^{-1} ~ @ ~100 ~\text{fb}^{-1}. 
\end{eqnarray}

Assuming the same selection as those used in 13 TeV analysis, 
prospects at 14 TeV at HL-LHC is obtained.
The constraints for $m_{a} = 1$ MeV at the center-of-mass energy of 14 TeV at 300 and 3000 fb$^{-1}$
are found to be:
\begin{eqnarray}
 \frac{c_{GG}}{f_{a}} \le   0.11 ~\text{TeV}^{-1}   ~ @ ~300 ~\text{fb}^{-1} ~,~ \frac{c_{GG}}{f_{a}} \le        0.063   ~\text{TeV}^{-1}  ~ @ ~3000 ~\text{fb}^{-1}.  
\end{eqnarray}

Regions in the ALP coupling $c_{GG}$ and mass at $95\%$ CL, 
from $t\bar{t}+a$ process are presented in Fig.\ref{fig:cgg-bound}.The regions are
corresponding to integrated luminosities of 2.2 and 3000 fb$^{-1}$.
As can be seen in Fig.\ref{fig:cgg-bound}, the result obtained from this channel 
is complementary to that derived from $j+\gamma+a$ process.
For the region $m_{a} \lesssim 0.04$ GeV, the $j+\gamma+a$ process
shows better sensitivity to $c_{GG}/f_{a}$  while in the mass region 
above $\sim 0.04$ GeV, $t\bar{t}+a$ channel provides stronger bounds.


\section{Constraining the ALP model by the top  quark dipole moments} \label{dipolet}

Several types of experiments are used to search for
the ALP model, ranging from the searches for direct production of ALP
at colliders to those from cosmological and astro-particle physics experiments. However, a complementary approach in hunting for ALP effects
would be the examination of the ALP indirect effects in higher order processes.
In this section, we derive upper limits on the
ALP coupling using the upper bounds on the top quark (chromo)magnetic dipole moment ((C)MDM).
Top quark dipole moments are useful observables
which enable us to constrain the parameter space
of new physics models \cite{Khatibi:2014bsa,Ayazi:2012bb}.
It is worth mentioning here that there are also studies to explain the 
anomalous magnetic moment of the muon by ($(g-2)_{\mu}/2$) 
employing a light pseudoscalar boson or ALP \cite{Chang:2000ii, Marciano:2016yhf,Bauer:2017nlg}.

\begin{figure}[h!]
\centering
\resizebox{0.65\textwidth}{!}{\includegraphics[width=1\textwidth]{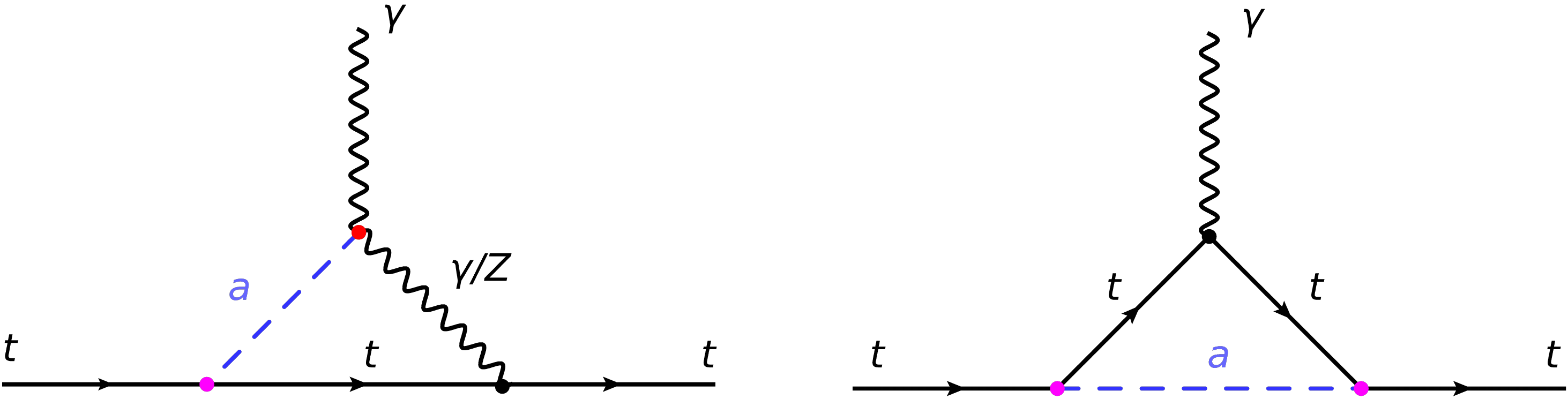}}\\
\resizebox{0.65\textwidth}{!}{\includegraphics[width=1\textwidth]{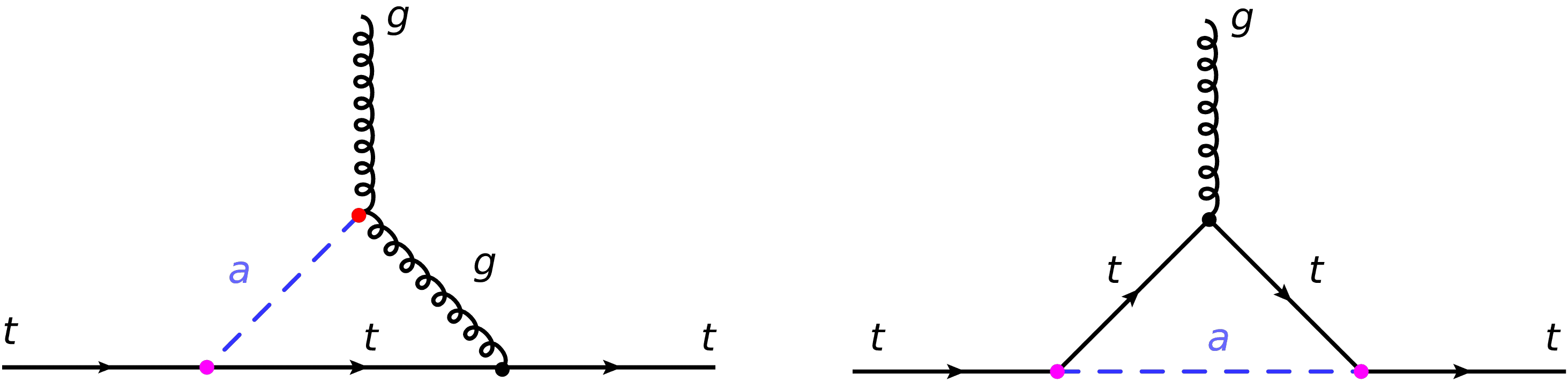}}
\caption{Feynman diagrams with contributions from an ALP to the top (chromo)electric and (chromo)magnetic dipole moments.}
\label{fig:diagram_dipole}
\end{figure}

The ALP interactions with photon, $Z$ boson,  gluons and top quark
as described in Lagrangian presented in Eq.\ref{bosonic_Lagrangian} lead to
considerable contributions  to the top quark dipole moments. 
The contributions from the ALP model to the top quark 
MDM are shown in the top plots demonstrated in Fig.\ref{fig:diagram_dipole}
and those contributing to the top quark CMDM are shown in the bottom 
plots in Fig. \ref{fig:diagram_dipole}.

It is notable that the dipole moments are proportional to 
two of the couplings of ALP at a time,  $c_{a\Phi}$ and $c_{GG}$ for chromo-dipole moments or 
$c_{WW(BB)}$ for electroweak dipole moments. Therefore, dipole moments provide
the possibility of probing two ALP couplings at a time.
The ALP model contributes to the top quark MDM ($a_{t}$) and CMDM ($\tilde{a}_{t}$) according to the
following relations:
\begin{eqnarray}
a_{t}&=& \frac{g_t - 2}{2} =  \frac{c_{a\Phi} \times c_{WW}}{f^{2}_{a}}  \frac{m_t^2 \sin^2\theta_{w}} { \pi^2 }  
  (\frac{3}{4}+\text{log}(\frac{\Lambda}{m_{a}})), \nonumber \\ 
 \tilde{a}_{t}&=& \frac{c_{a\Phi} \times c_{GG}}{f^{2}_{a}}  \frac{m_t^2} { \pi^2 }  
  (\frac{3}{4}+\text{log}(\frac{\Lambda}{m_{a}})).
\end{eqnarray}
As it can be seen, there is a logarithmic divergence to the dipole moments from the ALP model in photon (gluon) loops.
The contribution of the loop with $Z$ boson exchange is proportional to $\log(\Lambda^{2}/m^{2}_{Z})$ which is suppressed by the 
$Z$ boson mass therefore neglected here. 
The factor $\Lambda$
in the numerator of the logarithms which comes form the loop divergences
could be naturally assumed to be equal to the new physics scale.
In the ALP model, $\Lambda$ could be taken equal to $f_{a}$. Using the
upper limits on the top quark MDM and CMDM, one is able to constrain the 
ALP model parameters.

There are several studies to constrain the top quark dipole moments using 
the collider experiments and indirect searches which could be found in Refs.\cite{Bouzas:2012av, Fael:2013ira, CMS:2014bea,
Kamenik:2011dk, Etesami:2017ufk, Aguilar-Saavedra:2014iga , Etesami:2018mqk, Etesami:2016rwu}.
Limits on the top MDM obtained as $ -3 \leq a_t  \leq 0.45$ from the $t\bar{t} \gamma$ production at the Tevatron and the LHC.
The CMS experiment limits at $95\%$ CL on the CMDM of the 
top quark is $-0.043 \leq \tilde{a}_{t} \leq 0.117$~\cite{CMS:2014bea}.

\begin{figure}[h!]
\centering
	\resizebox{0.32\textwidth}{!}{\includegraphics[width=1\textwidth]{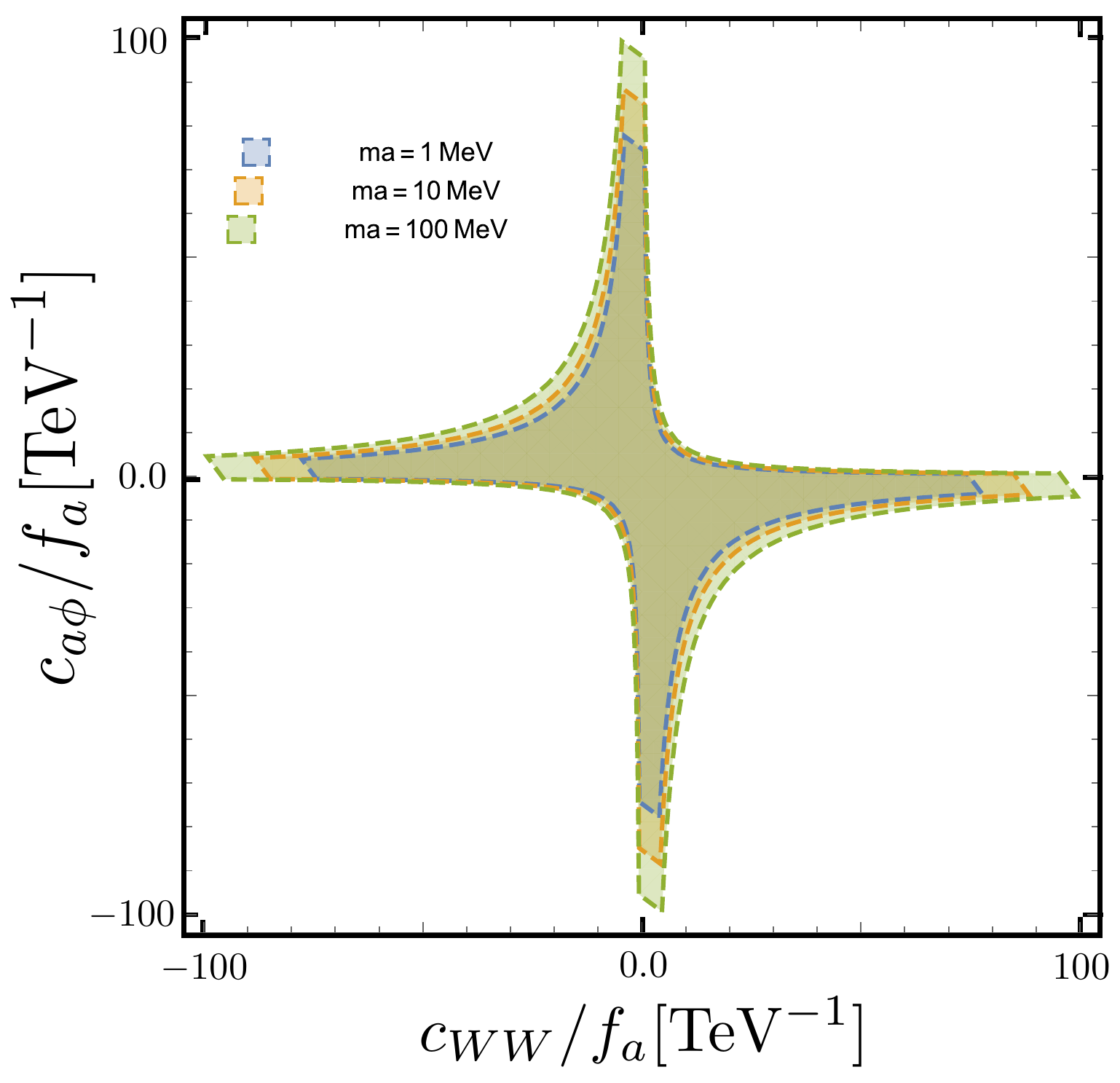}}
	\resizebox{0.32\textwidth}{!}{\includegraphics[width=1\textwidth]{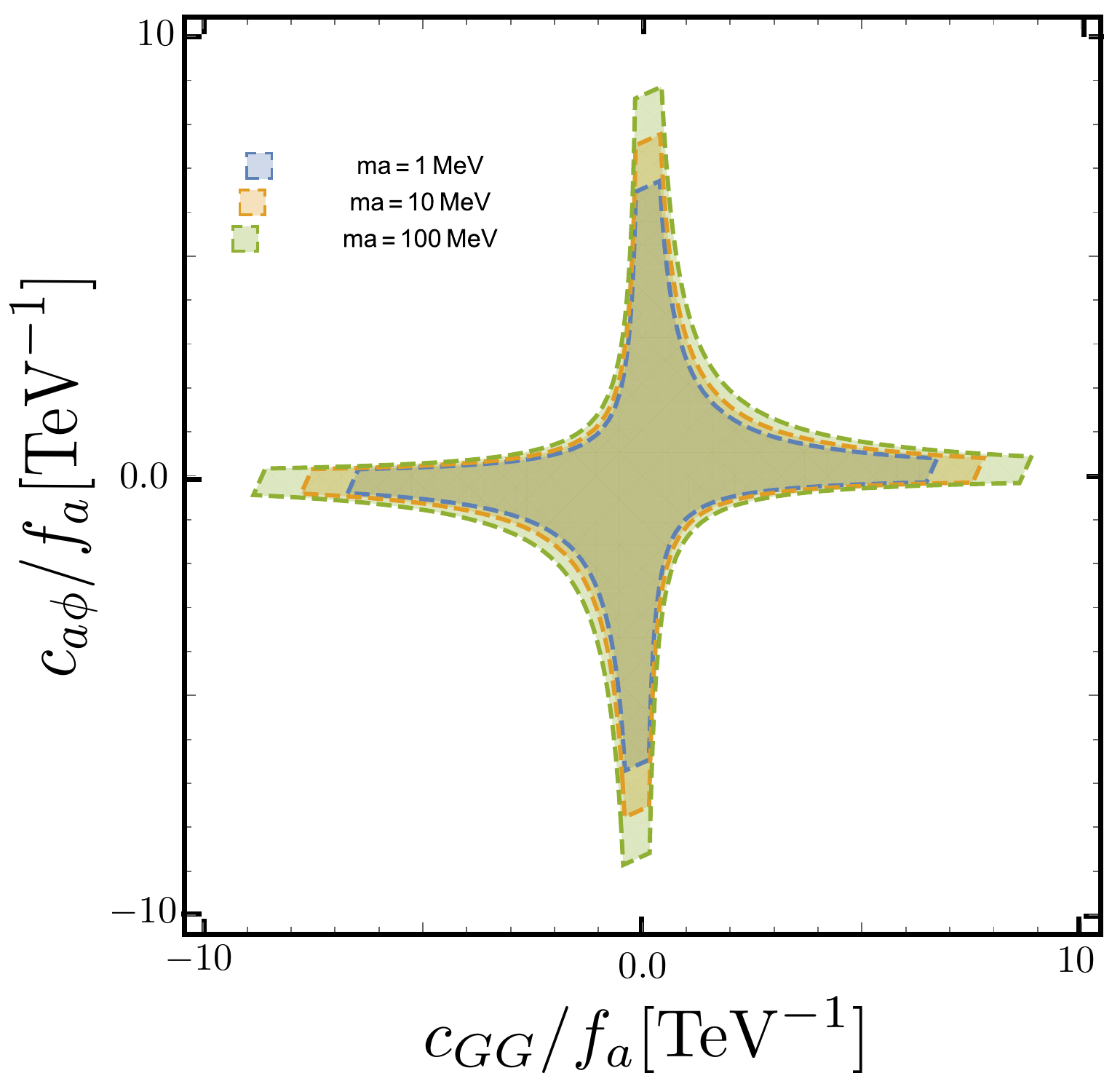}}
\caption{The $95\%$ CL exclusion limits on ($c_{a\Phi}/f_{a}$,$c_{WW}/f_{a}$) (left)   and 
($c_{a\Phi}/f_{a}$,$c_{GG}/f_{a}$) (right) using the top  quark MDM and CMDM limits.}
\label{fig:diagram_MDM}
\end{figure}

Figure \ref{fig:diagram_MDM} shows the $95\%$ CL exclusion regions on ($c_{a\Phi}/f_{a}$,$c_{WW}/f_{a}$) (left panel)   and 
($c_{a\Phi}/f_{a}$,$c_{GG}/f_{a}$) (right panel) obtained from the top quark MDM ($a_{t}$) and CMDM ($\tilde{a}_{t}$) limits, respectively.
The bounds are depicted for three values for the mass ALP $m_{a} = 1,10,100$ MeV. For smaller value of 
ALP mass, tighter bounds are achieved. 

The constraints are complementary to those derived from collider searches. In particular, 
for very light ALP masses, the magnetic dipole moments are powerful tools for exploration of 
the model parameters.

\section{ Summary and conclusions }\label{con}

Many SM extensions commonly predict the existence of light CP-odd scalars, so called axion like particles (ALPs),
which can couple to the SM gauge bosons and matter fields.  Such scalars  arise
from spontaneously broken global symmetries and provide
hints to answer some theoretical and observational defects of the SM
such as dark matter, baryogenesis, and strong CP problem.

In this study, to search for the new physics effects from ALPs, we consider the most general effective Lagrangian 
up to dimension five which describes the ALP interactions with SM fields.
We present collider limits on ALP model parameters from 
the $j+\gamma+a$, $t+j+a$, and $t\bar{t}+a$ search channels at the LHC with an integrated luminosity of
3000 fb$^{-1}$ of data.  It is found that the $j+\gamma+a$ channel is a promising process to 
probe both $c_{WW}$ and $c_{GG}$. In particular, the coupling of ALP with 
photon is reachable using this channel down to $|c_{WW}/f_{a}| \leq 0.036$ TeV$^{-1}$ for $m_{a} = 1$ MeV, which 
is at the order of the best expected limit from  mono-$Z$ process.  Reasonable sensitivity to 
the coupling of ALP to gluons $c_{GG}/f_{a} \leq 0.011$ TeV$^{-1}$ and $c_{GG}/f_{a} \leq 0.063$ TeV$^{-1}$ for $m_{a} = 1$ MeV
are obtained from the $j+\gamma+a$ and $t\bar{t}+a$ 
processes, respectively. Regions at the $95\%$ CL in the ALP parameter space 
$(c_{GG},m_{a})$ are found using $j+\gamma+a$ and $t\bar{t}+a$ channels 
which shows that the channels are complementary in probing the
ALP coupling to gluon.

\begin{figure}
\begin{center}
	\resizebox{0.62\textwidth}{!}{\includegraphics{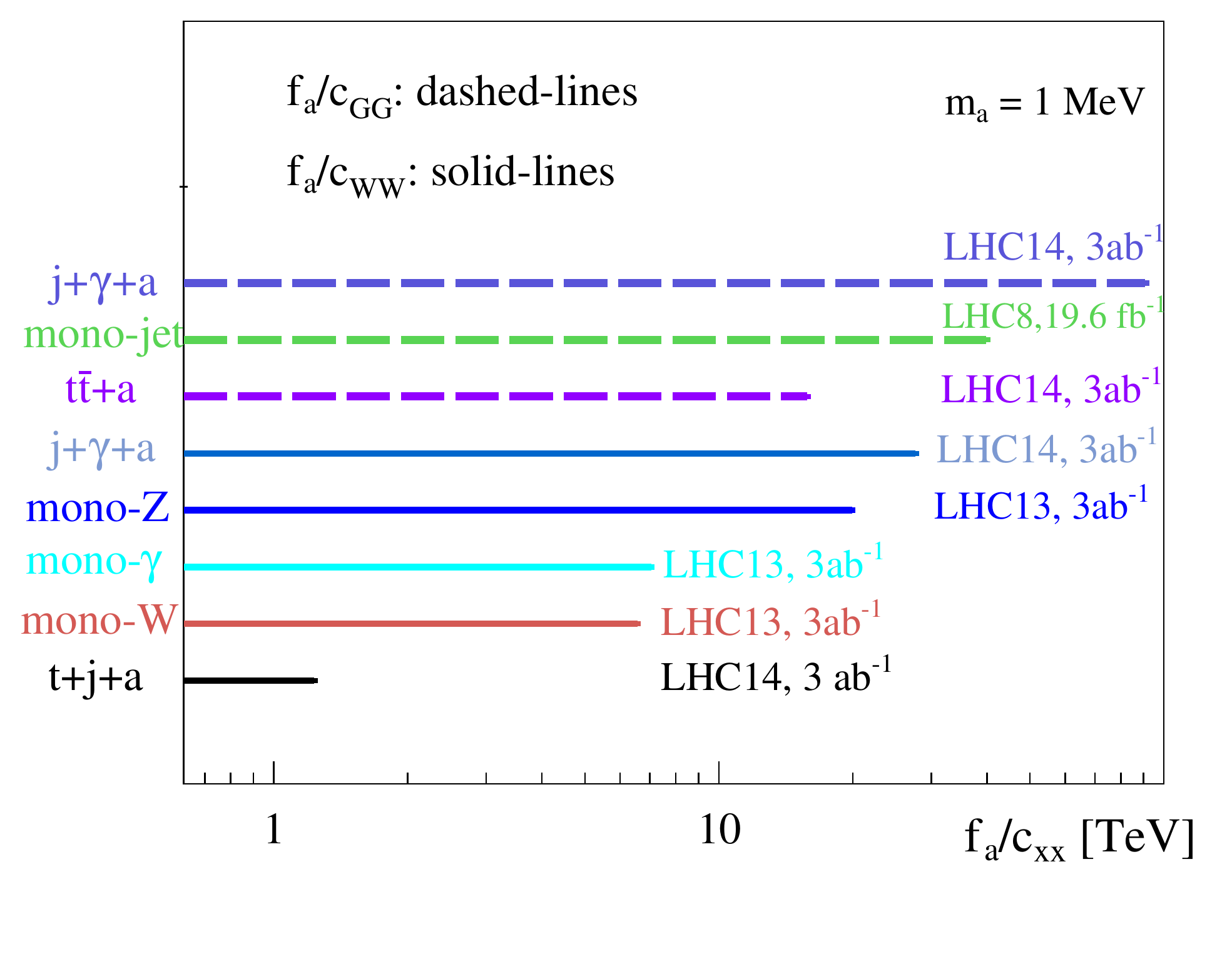}}
	\caption{The prospects for limits on $f_{a}/c_{GG}$ and $f_{a}/c_{WW}$ in TeV unit
	for $m_{a} = 1$ MeV at HL-LHC are
	presented. The results shown as $t+j+a$, top pair+$a$, and $j+\gamma+a$
      are derived from the current study. The other limits are taken from Refs.\cite{Brivio:2017ije,Mimasu:2014nea}. All limits
       are corresponding to an integrated luminosity of 3000 fb$^{-1}$ at $\sqrt{s} = 14$ TeV  except the one
       from mono-jet channel which obtained from 8 TeV LHC collisions data.}\label{fig:Limits}
	\end{center}
\end{figure}

In Fig.\ref{fig:Limits}, our prospect for bounds on the new physics energy scale from the ALP model over the
couplings $f_{a}/c_{GG}$ and $f_{a}/c_{WW}$ in TeV unit with $m_{a} = 1$ MeV are
presented. The bounds are the $95\%$ CL and correspond to 3000 fb$^{-1}$ of LHC
data at the center of mass energies of 13 TeV and 14 TeV.  The prospect for 
bounds from various channels are shown as well which are taken from Refs.\cite{Brivio:2017ije,Mimasu:2014nea}. 
All limits are corresponding to an integrated luminosity of 3000 fb$^{-1}$ at $\sqrt{s} = 14$ TeV except for
the one obtained from the mono-jet process which has been obtained from the LHC mono-jet 
analysis at 8 TeV collisions with 19.6 fb$^{-1}$ of integrated luminosity of data. 

At the end, we show that the 
ALPs could have a significant contributions to the top quark (chromo)magnetic dipole moment ((C)MDM).
Using the present limits on the top quark MDM and CMDM, 
the ALP couplings $c_{a\Phi}/f_{a},c_{WW}/f_{a}$ and $c_{a\Phi}/f_{a},c_{GG}/f_{a}$  
are constrained simultaneously. The exclusions regions are found to be tighter for lighter ALPs.

\section*{Acknowledgments}
We thank Andrea Giammanco for proposing data-driven ideas for background estimation,
reading the manuscript and giving several useful comments. The authors are grateful to Yotam Soreq
for the insightful comments and for providing data of  Ref. \cite{yotam}.
M.Mohammadi Najafabadi would like to thank INSF for the support.

%
%


%

\end{document}